# Enhanced chemical vapour deposition of monolayer $MoS_2$ films via a clean promoter

Lulin Wang[1,3], Yue Sun[1,3], Kaushik Kannan[1,3], Lee Gannon[1,3], Xuyun Guo[2,3], Aran Rafferty[2,3], Karl Gaff[1], Navaj B. Mullani[1,3], Haizhong Weng[1,3], Yangbo Zhou[4], Valeria Nicolosi[2,3], Cormac Mc Guinness[1,3], and Hongzhou Zhang[1,3,*]

[1]*School of Physics, Trinity College Dublin, Dublin2, Ireland*

[2]*School of Chemistry, Trinity College Dublin, Dublin 2, Ireland*

[3]*Centre for Research on Adaptive Nanostructures and Nanodevices (CRANN) and Advanced Materials and Bioengineering Research (AMBER) Research Centers, Trinity College Dublin, Dublin 2, Ireland*

[4]*School of Physics and Materials Science, Nanchang University, Nanchang 330031, China*

[*]*E-mail: hozhang@tcd.ie*

## Abstract

Two-dimensional (2D) transition metal dichalcogenides (TMDCs), exemplified by molybdenum disulfide ($MoS_2$), have shown exceptional potential for data-centred, energy-efficient electronic applications due to their unique electrical, optoelectronic, and mechanical properties. However, challenges such as the controllable synthesis of high-quality, large-area 2D $MoS_2$ films and the mitigation of contamination during growth remain significant barriers to their integration into advanced technologies. Here, we developed a novel contamination-free growth promoter, enabling the clean and scalable synthesis of high quality 2D $MoS_2$ with desirable grain structures via chemical vapour deposition (CVD). By optimising the reactant concentration and S/Mo ratio, we achieved promoter-dominated enhanced growth with enhanced quality, as evidenced by the increased MoS$_2$ flake size and coverage, alongside a strong PL A exciton peak at 1.84 eV, matching that of the mechanically exfoliated sample. This approach facilitates the clean and site-specific growth of high-quality 2D MoS$_2$, establishing a robust pathway for the practical implementation of 2D MoS$_2$ in next-generation electronic devices.

## Introduction

Two-dimensional (2D) transition metal dichalcogenides (TMDs) have emerged as promising candidates for next-generation electronic and optoelectronic devices, with emphasis on scaling and low-energy consumption applications, including logic transistor[1, 2, 3, 4, 5], photodetector[6, 7] and advanced neuromorphic devices[8, 9]. To ensure compatibility with industrial applications, it is crucial to achieve precise device control and minimize variability, which demands a controllable and scalable synthesis method for monolayer MoS$_2$.

Chemical vapor deposition (CVD) is recognized as one of the most promising approaches for scalable synthesis of MoS$_2$[10]. The epitaxial growth of MoS$_2$ on sapphire substrates has garnered considerable attention for achieving wafer-scale MoS$_2$ films[11, 12, 13, 14]. However, the as-synthesised MoS$_2$ requires an additional transfer step to relocate onto other arbitrary target substrates for further device fabrication, which may inevitably introduce contamination and strain that potentially alter or deteriorate the intrinsic properties of the film. Recently, the 2D Czochralski method[15] achieved uniform single-crystal MoS$_2$ with centimetre-scale domains on a molten glass substrate. While the etching-free transfer process is cleaner than the conventional wet transfer, it may still introduce strain, and the substrate introduced contamination (e.g. Na$^+$) remains unverified. Synthesizing MoS$_2$ directly on SiO$_2$/Si substrates provides a solution to these transfer and substrate related issues. However, the absence of epitaxial templating of MoS$_2$ on the amorphous SiO$_2$ surface makes it challenging to control the nucleation and large-scale synthesis of single-crystal MoS$_2$, resulting in random nucleation and other inhomogeneities[10]. Therefore, a clean approach for controlling MoS$_2$ nucleation and enhancing lateral film growth on amorphous SiO$_2$ substrate needs to be developed.

To facilitate the 2D growth of MoS$_2$, seeding promoters are introduced to the substrate prior to the CVD process. For instance, alkali metal halides (e.g. NaCl[16, 17], KI[16, 17]) and hydrates (e.g. NaOH[18]) have been reported to reduce the reaction temperature and weaken the interlayer adhesion, thereby enhancing the lateral growth of MoS$_2$ film. Similarly, wafer-scale 2D MoS$_2$ films are achieved when alkaline-earth metal chlorides (e.g. CaCl$_2$, SrCl$_2$[19]) are introduced in the CVD process. However, the residual metal ions from those promoters are detrimental to the electric properties of the MoS$_2$ film and it is found that compared to alkali-free synthesis under the same growth conditions, the utilization of NaCl can lead to variation in growth rates, loss of epitaxy, and dense clusters of nanoscale MoS$_2$ particles[20]. In addition, the alkali metal compounds lack the ability to control nucleation sites.

Alternatively, organic promoters such as perylene-3,4,9,10-tetracarboxylic acid tetrapotassium salt (PTAS) lead to millimetre size growth of 2D MoS$_2$, owing to the enhanced substrate adhesion of MoS$_2$ and heterogeneous nucleation sites[21, 22, 23]. However, uniformly applying PTAS on hydrophobic substrates poses challenges[23]. A thin film of polymer complex comprised of anhydrous ammonium tetrathiomolybdate (ATM) and linearpoly (ethylenimine) (L-PEI) has been found to effectively improve the growth uniformity and control the thickness of MoS$_2$ film in the range of 2-30 nm[24]. Nevertheless, precise control of the MoS$_2$ film thickness down to monolayer dimensions is still challenging. On the other hand, F$_{16}$CuPc has been shown to significantly enhance MoS$_2$ growth compared to other organic promoters due to its thermostability[23]. It is worth mentioning that non-volatile species, such as potassium in PTAS and copper in $F_{16}CuPc$, may be retained in the $MoS_2$ film, thus altering its intrinsic properties. Furthermore, the cleanliness of the film growth in previous reports is not thoroughly understood, and the absence of promoter-induced contamination is not yet well demonstrated.

Here, we developed a contamination-free method to realize the controlled growth of large-scale monolayer $MoS_2$ on an amorphous $SiO_2$ substrate. The growth, facilitated by nano promoters, yields $MoS_2$ free of contamination as confirmed by X-ray photoelectron spectroscopy (XPS) and Raman spectroscopy. The effect of the promoter is evaluated by modifying half of the substrate with nano promoters, whilst keeping all other growth parameters the same. A notable promoter effect is observed at a sulphur temperature of 250 °C, where the promoter modified region preferentially facilitates the growth of single-crystal monolayer $MoS_2$ films with over 3 times the coverage and a sixfold increase in the monolayer ratio. The proportion of large flakes (≥5 µm) found in the promoter region is 4 times that of the pristine region. Furthermore, we systematically investigated the promoter effect at varied reactant concentrations by tuning the sulphur temperature, demonstrating the transition from homogeneous nucleation to heterogeneous nucleation. In addition, we also found that the sulphur temperature greatly affects the sulphur vacancy density in $MoS_2$. Our results offer a novel strategy to achieve clean and controllable growth of large-scale 2D $MoS_2$ on amorphous $SiO_2$ substrates, establishing a solid groundwork for advancing the controllability and the performance of 2D electronic devices.

## Results

**The promoter effect: enhanced 2D growth**

The growth is achieved via the CVD method with the assistance of polymer nano promoters. Specifically, the promoter is a conventional photoresist S1813 (Shipley), which consists of cresol novolak resin and a photoactive compound (PAC)[25]. To investigate the promoter effects, the substrate surface is divided into two regions; one decorated by the promoter, the other kept pristine (Fig. S1). In a typical growth process (see Fig. 1a top panel), a quartz crucible is loaded with 1 mg $MoO_3$ power (Afla Aesar, 99.9%). The substrate is positioned 0.5 cm downstream from the $MoO_3$ powder and tilted at an angle of 120° relative to the tube axis. Additionally, 120 mg of sulphur powder is placed in a separate crucible. Both crucibles are loaded into the furnace with each crucible's temperature controlled independently. The heating controller of sulphur is then energised when the substrate temperature reaches 765 °C. During the growth process, the substrate/$MoO_3$ temperature is maintained at 835 °C, and the sulphur temperature is kept at 250 °C for 5 minutes. Nitrogen gas is introduced as a protective and carrier gas at a flow rate of 20 sccm throughout the reaction period. Detailed information on substrate preparation and the CVD process can be found in the Methods Section.

We observe that the distance between the molybdenum source and the deposition site on the substrate can affect the growth behaviour due to variations in gaseous reactant concentration and flow. As such, for a set of growth parameters, we compare the growth behaviours at two equivalent locations sampled from the pristine and promoter regions of the substrate, which are subject to an equivalent reaction environment. The most significant growth enhancement is observed by comparing two such equivalent regions near the bottom of the substrate and close to the crucible side walls. These regions, indicated by the dashed circles in the bottom panel of Fig. 1a, are located 12 mm from the molybdenum source. The promoter region (right) exhibits a distinctly different colour compared with the pristine region (left). A continuous dark olive-green film is visible on the surface of the promoter region, while the contrast of the pristine surface is close to that of the bare $SiO_2$/Si substrate. The difference in optical contrast suggests that, at the macroscopic scale, the promoter region has significantly more growth than the pristine region.

The optical micrographs of the pristine region (Fig. 1b) and promoter region (Fig. 1c) corroborate the observations made through visual inspection. Sparse micrometre-sized particles are present in the pristine region with diminished 2D growth. In contrast, the promoter region features a uniform distribution of triangular $MoS_2$ flakes. Some triangular flakes merge, forming films up to approximately 33 μm in size. Optical micrographs with a field of view of 500 μm × 500 μm were further used for statistical analysis, which is conducted using Python[26]. Only flakes larger than 2 μm are counted, taking into account the resolution of the optical microscope. Approximately 534 small particles are identified in the pristine region, compared with about 224 µm-sized flakes in the promoter region (Fig. S2a-b). The histogram in Fig. 1d illustrates that the flake size distribution in the pristine region follows a lognormal pattern with an average size of 3.9 μm (flake recognition see Fig. S2a). The promoter area exhibits a bimodal Gaussian distribution, where the first peak at 6.4 μm corresponds to the small flakes (as marked by white arrows in Fig. 1c and shown in Fig. S2c), which overlaps with the peak of the pristine region, indicating homogenous nucleation. The main peak at 14.4 μm represents the large triangular flakes (Fig. S2d), which is more than 3 times larger than the pristine region, indicating the heterogeneous nucleation induced by the nano promoters. Figure 1e compares five key characteristics of CVD growth, including the average size. The monolayer ratio, defined as the area ratio of monolayer $MoS_2$ to the total recognized area (Fig. S2a-b), of the promoter region is 88.9%, approximately six times higher than that of the pristine region. The abundance percentage of large flakes (>10 µm) is 30.2% in the promoter region and 0.5% in the pristine region. The coverage, defined as the ratio of the area covered by flakes to the total substrate area in the image, is significantly higher in the promoter region at 15.0 %, almost four times that of the pristine region (4.0 %). However, the population density, representing the number of flakes or particles per unit area, is lower in the promoter region ($1.0 \times 10^3$ mm$^{-2}$) than in the pristine region ($2.4 \times 10^3$ mm$^{-2}$). These quantitative statistical results confirm the enhanced growth of 2D MoS$_2$ in the promoter region.

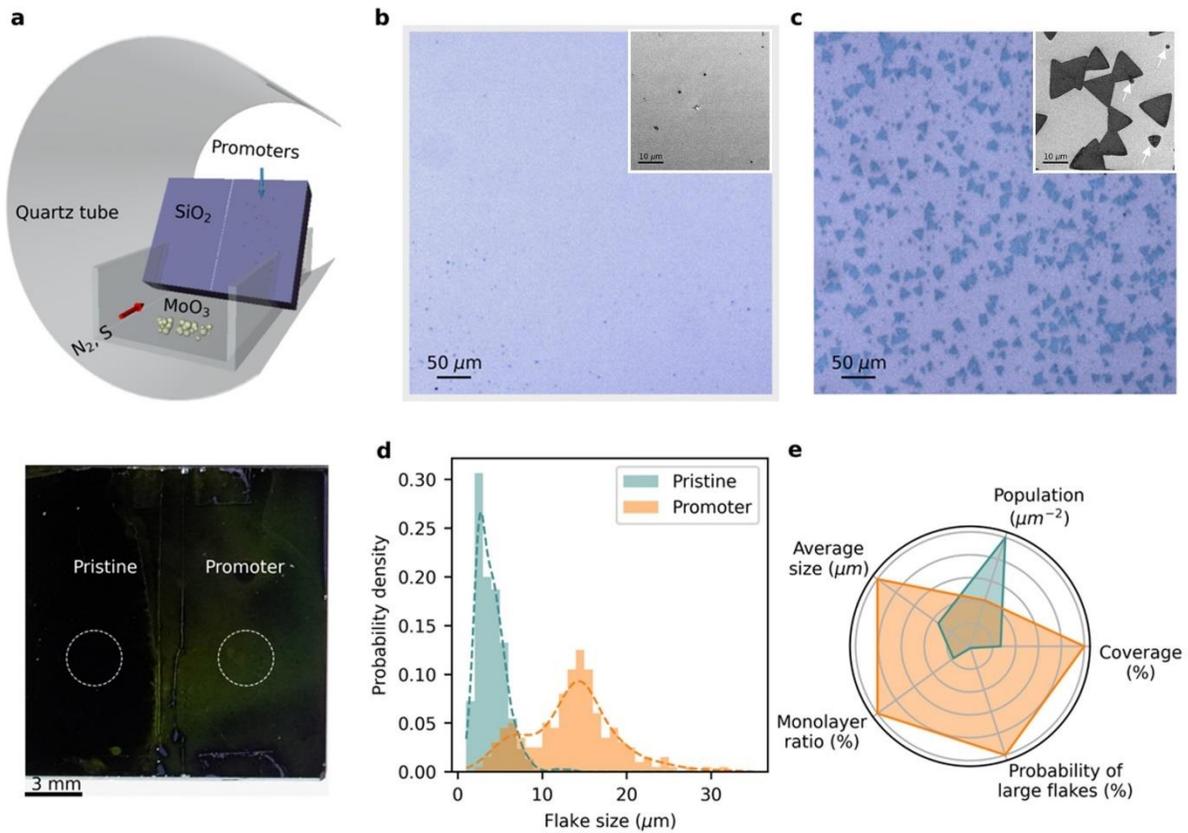

Figure 1: A typical promoter-assisted CVD growth of MoS$_2$ at a sulphur temperature of 250 °C. (a) Top panel: an illustration of the key growth configuration. Bottom panel: An optical image of the wafer coupon post-growth. The white dashed circles highlight two equivalent regions—one from the pristine region and the other from the promoter region—under similar growth conditions, such as reactant concentration and flow. Optical micrographs for (b) the pristine substrate which remain clean and (c) the promoter region covered by triangular MoS$_2$ flakes. (d) Statistical distribution of flake sizes comparing the growth in the pristine and promoter regions. (e) A radar chart comparing the average flake size, monolayer ratio, probability of large flakes, coverage and population density for the pristine substrate and promoter regions.

The MoS$_2$ flakes in the promoter region are further characterized by Raman spectroscopy and photoluminescence (PL). As shown in Fig. 2a, the average Raman spectrum exhibits two significant peaks at 384.3 cm$^{-1}$ and 405.1 cm$^{-1}$ which correspond to the E$^1_{2g}$ and A$_{1g}$ modes of 2H MoS$_2$, respectively. The spacing, $\Delta\omega$, between the two Raman modes is 20.8 cm$^{-1}$ whereas the peak intensity ratio of E$^1_{2g}$ and A$_{1g}$ is 0.72. These results again indicate the monolayer nature of the MoS$_2$ flake[27, 28, 29, 30]. Additionally, from the full-range Raman spectrum (Fig. S3a), no peak is observed in the Raman spectrum range of 1100 to 2000 cm$^{-1}$, indicating the absence of carbon residue[31]. A weak LA peak can be observed in the range of 210 to 250 cm$^{-1}$, which is a typical Raman mode contributing to defects[32]. A map of the $\Delta\omega$ is shown in Fig. 2b. The majority (~95%) of the flakes in the mapping region exhibit consistent values of $\Delta\omega \sim 20$ cm$^{-1}$, suggesting a high population density of monolayer flakes. However, around 5% of areas show $\Delta\omega \sim 22$ cm$^{-1}$, which is attributed to bi-layer MoS$_2$. The monolayer ratio is consistent with the optical contrast result in Fig. S3b. Additionally, secondary nucleation is evident in the scanning electron microscopy (SEM) image shown in Fig. S3c. In the PL spectrum (Fig. 2c), a strong luminescence band appears at 1.845 eV and corresponds to the direct excitonic transition energy (A exciton)[2], indicating the direct band gap of the monolayer MoS$_2$[3]. Moreover, the B exciton resonance is observed at 1.97[3]. The thickness of the flake is further verified through atomic force microscopy (AFM) measurement as shown in Fig. S3d. The line profile in Fig. S3e indicates that the MoS$_2$ film thickness is 0.68 nm.

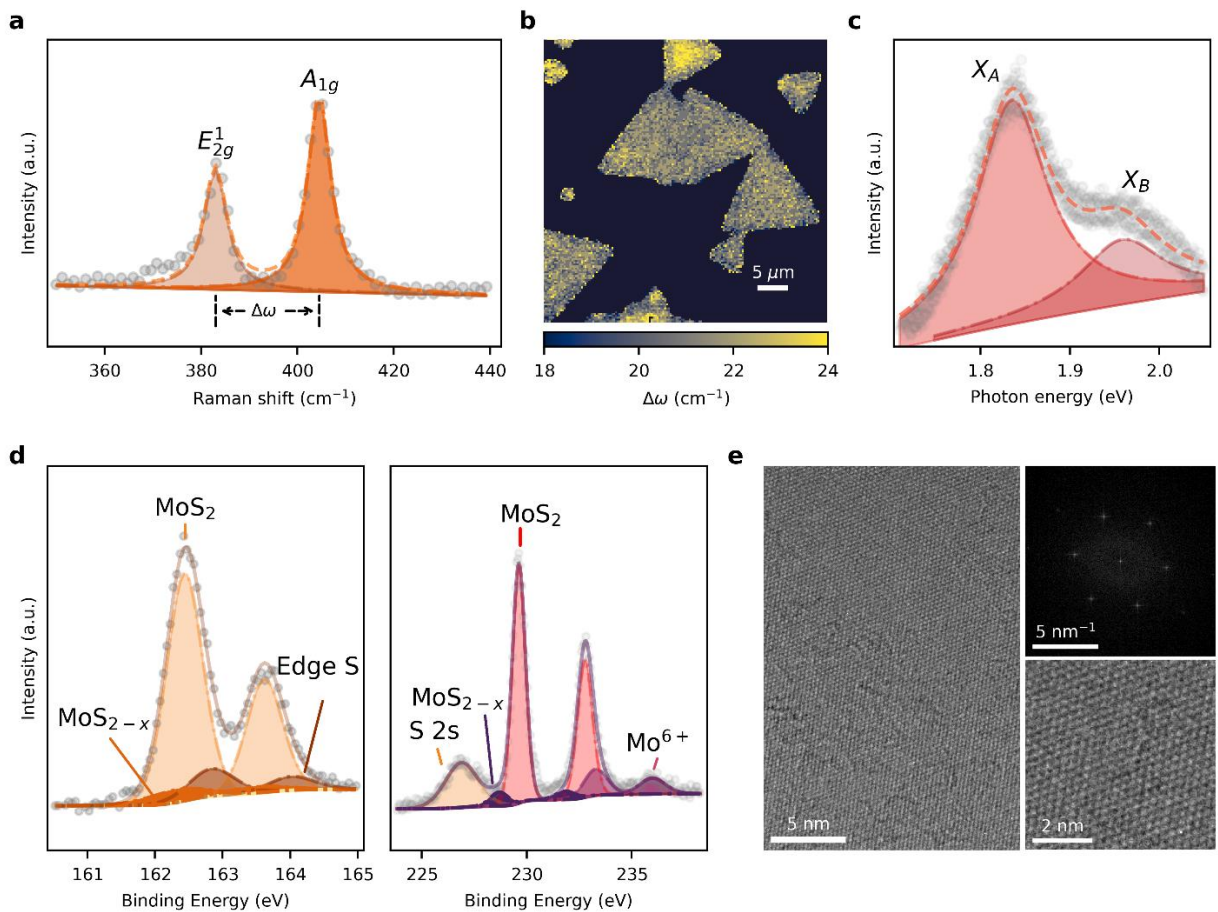

Figure 2: Characterisations of the MoS$_2$ flakes in the promoter region under sulphur temperature of 250 °C. (a) The average Raman spectrum shows E$^1_{2g}$ and A$_{1g}$ modes of 2H MoS$_2$ at 384.3 cm$^{-1}$ and 405.1 cm$^{-1}$ with a peak spacing of 20.8 cm$^{-1}$. (b) The mapping of the peak position difference shows the high population of monolayer flakes. (c) Averaged PL spectrum shows strong A exciton at 1.845 eV indicating the direct band gap MoS$_2$ which corresponds to the behaviour of monolayer MoS$_2$. (d) XPS core-level spectra of Mo 3$d$ and S 2$p$, where the main peaks show the chemical bond in 2H MoS$_2$. (e) HRTEM shows the lattice planes of MoS$_2$. The top right panel is the Fourier transform of the HRTEM image, which implies a single crystalline hexagonal structure. The zoom-in HRTEM image in the bottom right panel confirms the disorder-free lattice and the 2H phase.

Figure. 2d shows the XPS of core level spectra S 2$p$ and Mo 3$d$ spectra of the films in the promoter region. In the S 2$p$ core-level spectrum, the doublets at 162.4 eV and 163.6 eV indicate the chemical states of S in 2H MoS$_2$[4, 33]. The weak peaks at 164.1 eV and 162.9 eV can be attributed to the edge sulphur[34]. The doublet at 162.3 eV and 163.5 eV corresponds to the reduced chemical state of S, indicating a MoS$_{2-x}$ phase. In the Mo 3$d$ core level spectrum, the doublets at 229.6 eV and 232.8 eV correspond to the chemical states of Mo in 2H MoS$_2$[4, 33]. The relative weak peaks at 233.3 eV and 236.0 eV correspond to the Mo$^{6+}$, implying the presence of a small amount of MoO$_3$. Doublet at 228.70 eV and 231.84 eV corresponds to the MoS$_{2-x}$ phase. The peak at 227.0 eV is attributed to the S 2$s$. The results suggest that the film primarily consists of 2H MoS$_2$, with detectable sulphur vacancies and oxides species. For completeness, we investigate the particles grown in the pristine region. As shown in Fig. S4a and S4b, the XPS spectra of the pristine region show a weak signal in the range of 224 –239 eV (Mo 3$d$ core level), while no peak is detected within the binding energy range of 162–164 eV (S 2$p$ core level). This suggests the presence of a small amount of molybdenum compound but no detectable MoS$_2$. Moreover, Fig. S4c and S4d show the morphology of the pristine particles and the corresponding Raman spectrum. The characteristic peaks at 203, 233, 363, 464, 496, 570, and 744 cm$^{-1}$ in Fig. S4d suggest that the particles are crystallized in the MoO$_2$ phase[35, 36]. The formation of MoO$_2$ rather than MoS$_2$ suggests an ineffective sulphurisation reaction in the pristine region. The high-resolution transmission electron microscopy (HRTEM) image, see Fig. 2e, shows the lattice of 2H MoS$_2$. The six-fold symmetry is evident in the Fourier transform of the image shown in the top right panel of Fig. 2e. The 2D lattice shown in the bottom right panel confirms the high-quality lattice structure and corroborates the 2H phase. The optical properties, chemical state and the lattice structure confirm that the flakes are monolayer 2H MoS$_2$.

To verify the cleanliness of the promoter-enhanced CVD method, XPS C 1s and O 1s core-level spectra are analysed and compared in Figure S5. For both the pristine and promoter regions, in the O 1s core-level spectra shown in Fig. S5a-S5b, the main peak at 533.2 eV represents the O-Si bond which arises from the substrate. The weak peaks at 532.2 eV and 531.0 eV correspond to O-C and O-Mo bonds[37]. In the C 1s core-level spectra (Fig. S5c-S5d), a prominent peak appears at approximately 285.0 eV for both the pristine and promoter regions, corresponding to C-C bonds. In addition, peaks at around 286.5 eV, 288.0 eV and 289.15 eV correspond to the C-O single bonds, C=O double bonds and O-C=O bonds, respectively. These peaks are typical for adventitious carbon[38, 39] which is usually found on the surface of most air-exposed samples. There is no evidence to show that C and O bond with Mo or S (for example, C-Mo bonds in C 1s at 284.2 eV[39, 40], C-S-C in S2p at 163.5 eV and 164.8 eV[41]). The adventitious C signal is stronger in the promoter region than in the pristine region. A comparison of C 1s core-level spectra of the bare $SiO_2$ substrate and the bulk $MoS_2$ (Fig. S5e-S5f) shows consistently higher C intensity on $MoS_2$. This implies stronger C adsorption on the $MoS_2$ surface than the $SiO_2$ possibly due to the lower surface energy of $MoS_2$ relative to $SiO_2$[42, 43]. The results confirm the cleanliness of the nano promoter-assisted growth.

**Effects of the sulphur-source temperature**

To further investigate the effect of the promoter, we examined the evolution of the morphology and yield of $MoS_2$ as a function of the sulphur temperature. The morphology of the CVD products (Fig. 3a) in the promoter region varies depending on the sulphur temperature. When the temperature is 180 °C, hexagonal $MoS_2$ flakes with rounded edges and a size of approximately 3 μm are observed, indicating insufficient sulphur concentration. As Ts increases to 200 °C, the $MoS_2$ flakes transition to a triangular shape with soft edges, suggesting an increase in sulphur concentration though it remains inadequate[44, 45]. At 220 °C, the triangular flakes grow larger and exhibit straighter edges. Further increasing to 250 °C results in uniformly-distributed triangular

MoS$_2$ flakes with increased flake size and well-defined straight edges. This indicates that the reactant concentration is sufficient for the 2D growth of stoichiometric MoS$_2$. For a temperature of 270 °C, the flakes show irregular shapes with sizes ranging from 0-30 μm, indicating the merging of multiple grains. The darker contrast at their centre suggests excessive growth of multilayer films, accompanied by the appearance of particles. Finally, at 300 °C, the MoS$_2$ appears as particles and submicron-sized triangular flakes. The presence of particles rather than the formation of films suggests that the nucleation rate has exceeded the growth rate.

Further statistical analysis, derived from the optical images in Fig. S6 and summarised in Fig. 3b, compares five key growth metrics in the pristine and promoter regions. Multiple data points at a given temperature represent distinct samples grown under identical parameters, with error bars indicating the variation among flakes within each sample. The average flake size in the pristine region remains consistently within the range of 3–5 μm across all temperatures. In the promoter region, the average flake size is approximately 4 μm at 180 °C and increases with the sulphur temperature, reaching a peak of 9.8 μm at 250 °C and 11.2 μm at 270 °C. However, beyond 270 °C, the average flake size decreases, dropping to around 4 μm at 300 °C. The MoS$_2$ flakes in the promoter region are consistently larger than those in the pristine region, with the size difference peaking between 250 °C and 270 °C.

The monolayer ratio, in the pristine region, remains around 5% between 180 °C and 200 °C, reaching a maximum of 20% at 220 °C and 250 °C, and decreasing to 7% at 300 °C. In the promoter region, the monolayer ratio increases steadily with temperature, peaking at 82% at 250 °C, before declining to 48% at 300 °C. The promoter region exhibits a higher monolayer ratio at all temperatures, with the maximum difference between the two regions observed at 250 °C.

The probability of large flakes, on the pristine side, remains steady at around 1.0% between 180 °C and 300 °C. On the promoter side, the probability is 0% at 180 °C, increases to 12.6% at 200 °C, rises significantly to a 35%-50% between 220 to 270 °C, and then decreases to 3.4% at 300 °C. These results show that the promoter region has a higher probability of large flakes, with the maximum occurring between 220 to 270 °C.

The promoter region exhibits significantly higher flake coverage compared to the pristine region within the sulphur temperature range of 200 °C - 270 °C, peaking at 250 °C (18%). In contrast, coverage in the pristine region remains relatively low between 0.5-2% across the temperature range of 180 °C - 300 °C. This highlights the enhancement of $MoS_2$ growth in the promoter region, particularly at a temperature of 250 °C.

The population density, on the pristine side, increases from $0.8\times10^3$ to $3.1\times10^3$ $mm^{-2}$ as the temperature rises from 180 °C to 200 °C. However, as the temperature continues to increase, it decreases to $1.3\times10^3$ $mm^{-2}$ at 220 °C and remains around this value between 220 °C to 300 °C. On the promoter side, except at 180 °C, the population density is lower than that of the pristine side. At 180 °C, the population density on the promoter side is 2.5%, whereas between 200 °C and 300 °C, it ranges from 1.1% to 2.2%.

From the statistical results, the small average size and high population density indicate a high nucleation rate and a low film growth rate in the pristine region. It is evident that $MoS_2$ 2D growth in the promoter region is significantly enhanced, particularly for temperatures between 220 °C and 270 °C, as indicated by increased flake sizes, monolayer ratio, probability of large flakes, and coverage.

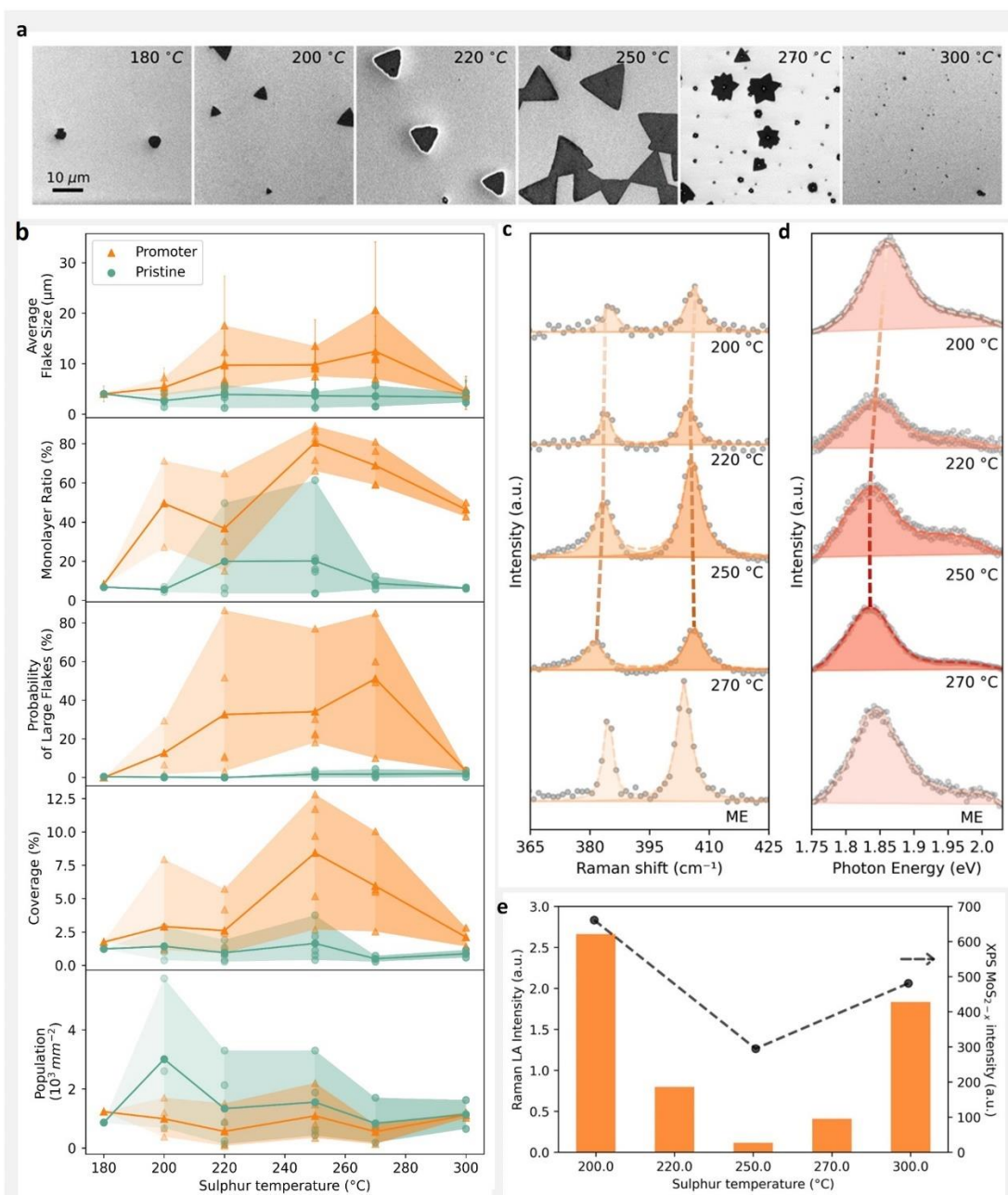

Figure 3: MoS$_2$ properties at varied sulphur temperature. (a) SEM images of samples in the promoter region. (b) Statistic results of MoS$_2$ average flake size, monolayer ratio, probability of large flakes, coverage and population density, which are based on large-area optical images shown in Fig. S5. (c) Average Raman spectra of CVD MoS$_2$ and ME monolayer MoS$_2$. The spectra are normalized against the Si peak at 520 cm$^{-1}$ and fitted by Lorentzian function. (d) Average PL spectra of CVD MoS$_2$ and ME monolayer MoS$_2$. (e) Defect analysis of 200 °C, 250 °C and 300 °C samples by Raman LA peak intensity and XPS MoS$_{2-x}$ phase intensity.

To further evaluate the quality of the synthesized MoS$_2$ under different sulphur temperatures, Raman and PL characterizations are systematically conducted. The average Raman spectra in Fig. 3c obtained from the Raman mappings (Fig. S7), which are normalized against the Si peak at 520 cm$^{-1}$ and fitted by Lorentzian functions. The normalisation removes intensity variations caused by fluctuations in laser power. No detectable Raman peaks are observed at 180 °C, which suggests a limited amount of deposition. The Raman spectra for sulphur temperature between 200 °C to 270 °C exhibit the characteristic A$_{1g}$ and E$^1_{2g}$ modes of MoS$_2$. For temperatures between 200 °C and 250 °C, the E$^1_{2g}$ is at 383.8±0.7 cm$^{-1}$, while it redshifts to 381.9 cm$^{-1}$ at 270 °C. The A$_{1g}$ peak is at 406.4 cm$^{-1}$ at 200 °C, which then features a redshift to ~405.2 cm$^{-1}$ at higher sulphur temperatures. The peak difference, Δω, varies accordingly. From 200 °C to 220 °C, the peak difference is around 22.1 cm$^{-1}$. At 250 °C, the peak difference is 20.8 cm$^{-1}$ which is the minimum among all the samples, indicating the monolayer thickness. When the sulphur temperature exceeds 250 °C, Δω increases, reaching 24.4 cm$^{-1}$ at 300 °C. The increase in Δω indicates an increase in MoS$_2$ thickness[46]. Notably, the E$^1_{2g}$ and A$_{1g}$ peak positions of the sample grown at 250 °C are comparable to those of the ME sample, suggesting that the film quality is close to the ME sample (Fig. S8).

The average PL spectra in Fig. 3d reveal variations in the Xa exciton energy across different samples. No signal is observed at 180 °C and 300 °C, indicating weak light emission which may be due to the indirect bandgap. For MoS$_2$ synthesized at 200 °C, the Xa peak appears at 1.876 eV. As the temperature increases from 220 °C to 250 °C, the Xa peak redshifts to 1.851 eV and 1.845 eV (consistent with the ME sample), respectively. At 270 °C, the Xa peak shows a slight blueshift to approximately 1.846 eV. Compared to 250 °C, the blueshift of 0.031 eV observed at 200 °C suggests that excessive oxygen is incorporated into the film under a sulphur deficient condition, which also enhances the PL intensity[47]. The PL intensity is weak at 220 °C increases at 250 °C and decreases again at 270 °C. These results indicate that the highest film quality is achieved at 250 °C.

In addition to the main Raman vibration modes of $MoS_2$ shown in Fig. 3c, the longitudinal acoustic (LA) mode in the frequency region 210 - 250 cm$^{-1}$ (Fig. S9) appears exclusively for the defective structure, and is a distinct signature of the S vacancy[32]. The integrated intensity of the LA peaks, shown in Fig 3e, decreases as sulphur temperature increases from 200 - 250 °C and rises when the temperature exceeds 250 °C, indicating the lowest defect concentration in the sample at a temperature of 250 °C. On the contrary, the sample at a temperature of 200 °C exhibits the highest LA intensity. The strong disorder-induced scattering reflects a higher defect density. Additionally, XPS is employed to characterize the temperature dependence of the sulphur vacancies (Fig S10). The integrated intensity of the non-stoichiometric phase $MoS_{2-x}$ is the lowest for the sample at the temperature of 250 °C, which corroborates the Raman defect analysis. The sulphur vacancies in the 200 °C sample are further validated through TEM analysis, as shown in the ACSTEM images (Fig. S11).

## Discussion

To explore and understand the significance of promoter and sulphur in the growth process of $MoS_2$, thermo-gravimetric analysis (TGA) is conducted to compare the thermal behaviour of the $MoO_3$ with and without sulphur, as shown in Fig. 4a. The Stage 1, ranging from room temperature to 330 °C, corresponds to sulphur sublimation, while Stage 2 spans from 330 °C to 740 °C, preceding the onset of $MoO_3$ sublimation. For $MoO_3$ without sulphur, there is no weight loss during Stage 1 and the weight loss in Stage 2 is 0.83%. In contrast, for the $MoO_3$ with sulphur, the initial sulphur to total mass ratio is 30.79%. The weight loss of Stage 1 is 28.79% indicating residual sulphur that may form $MoO_{2-x}S_x$. During Stage 2, the weight loss is 1.96%, giving the total weight loss of 30.75% after Stage 2, suggesting that sulphur remains in the compound.

The differential thermo-gravimetric (DTG), or first derivative of the weight with respect to time, curves plotted in Fig. 4b shows that the maximum sublimation rate of $MoO_3$ is higher in the presence of sulphur (0.053 mg/s) compared to its absence (0.046 mg/s). Furthermore, $MoO_3$ with sulphur reaches its maximum sublimation rate earlier than without sulphur, indicating an acceleration of sublimation. This increased sublimation rate may result from the formation of intermediate product $MoO_{2-x}S_x$, which has weaker Mo-S bonds compared to Mo-O bonds[48], facilitating decomposition and sublimation. These findings suggest that the $MoO_3$ sublimation rate and gas phase concentration can be tuned by the sulphur concentration. In our experiment, the sulphur temperature plays a crucial role in regulating the sulphur evaporation rate. A higher sulphur temperature results in a greater sulphur concentration in the ambient, leading to an increased S-to-Mo ratio.

The sulphur concentration depends not only on the temperature but also on the presence of nano promoters. The sulphur in the nano promoters (Fig. S12) locally increases the fractional sulphur concentration, promoting the reduction and sulphurisation of $MoO_3$, thereby facilitating $MoS_2$ nucleation. In contrast, in the pristine region, the insufficient reduction of $MoO_3$ results in the formation of $MoO_x$ instead of $MoS_2$. Furthermore, the decomposition of nano promoters at high temperatures releases carbon, which further enhances the reduction of $MoO_3$ and inhibits the formation of $MoO_x$.

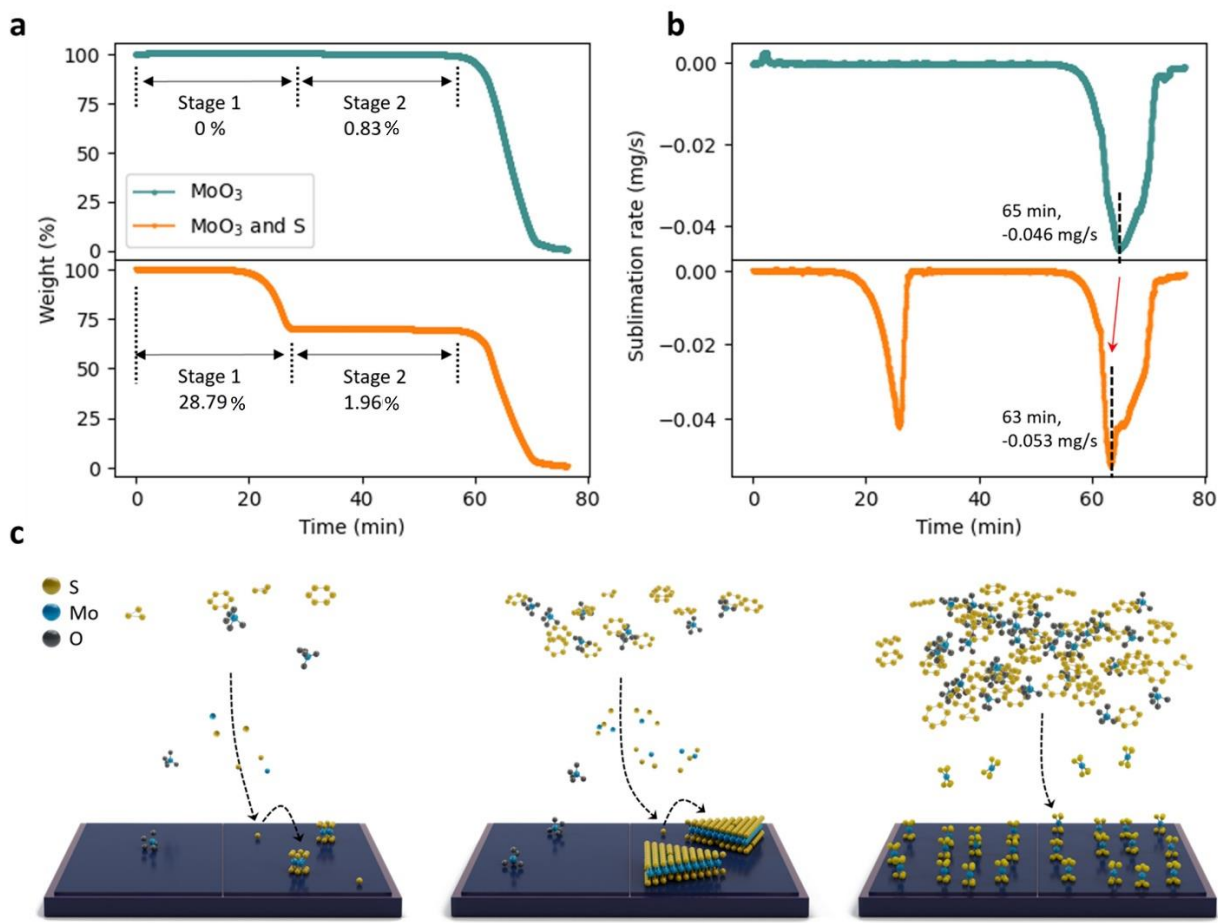

Figure 4: (a) TGA and (b) DTG curves of $MoO_3$ with and without sulphur. (c) A schematic illustration of the promoter effects at sulphur temperature of 200 °C (left), 250 °C (middle) and 300 °C (right).

Figure 4c presents a schematic illustrating the effects of the promoter at varying sulphur temperatures. At low temperature (e.g. 200 °C), the ambient sulphur concentration is insufficient. In the pristine region, $MoS_2$ nucleation is hindered due to the sulphur deficiency. In contrast, in the promoter region, sulphur supplied by the nano promoter enables $MoS_2$ nucleation. However, the overall low sulphur concentration remains inadequate to support extensive film growth. Consequently, the resulting flakes are small and exhibit either a hexagonal geometry or small triangles with rounded edges, which are likely to contain significant sulphur vacancies.

The sulphur concentration increases as the temperature rises to 250 °C. In the promoter region, a consistent and moderate supply of reactants facilitates the 2D film growth with a triangular geometry and fewer defects. Additionally, as previously discussed, the decomposition of the nano promoter at high temperatures provides carbon, which facilitates the reduction of $MoO_3$ and suppresses the deposition of $MoO_x$ on the substrate. This enhanced reduction of Mo combined with a sufficient supply of reactants, promotes preferential growth in the promoter region.

At higher temperatures (e.g. 300 °C), there is abundant sulphur in the ambient, which also accelerates the consumption of $MoO_3$, leading to a peak concentration of both S and Mo reactant within a short timeframe. The supersaturated reactant concentration leads to a mass transport-limited CVD process[49], resulting in a high density of particles across the entire substrate, regardless of promoter treatment. This indicates a highly homogeneous nucleation process in both the pristine and promoter regions. Excess reactants spontaneously nucleate numerous $MoS_2$ particles throughout the substrate, without influence from any foreign surface (e.g. promoters). It's important to note that homogeneous nucleation doesn't exhibit any preferential attachment to the nano promoters on the substrate, rather it results in a more dispersed distribution of $MoS_2$ particles. During the nucleation stage, the rapid depletion of S and Mo reactants limits the availability of reactants for further film development.

From our experiments, we found that heterogenous nucleation dominated $MoS_2$ growth occurs when the sulphur temperature is between 200 °C and 300 °C. This process is regulated by controllable nanoscale promoters in terms of position and density, leading to a more organized and well-structured growth compared to random homogeneous nucleation.

# Conclusion

In summary, we employed nanosized polymer particles as promoters to control the $MoS_2$ nucleation process, transforming it from homogenous to heterogenous. By controlling the sulphur temperature, the sulphur partial pressure and the reactant concentration can be tuned. We demonstrate the pivotal role of nanosize polymer promoters in enabling heterogeneous nucleation while suppressing homogeneous nucleation during the CVD growth of $MoS_2$. Our findings show that a supersaturated reactant concentration result in excessive homogenous nucleation, regardless of the presence of a polymer promoter, while moderate reactant concentrations facilitate promoter-assisted heterogenous nucleation. The film quality is comprehensively investigated using optical imagery, SEM, PL, AFM, and Raman for the condition of varied reactant concentrations. We found that $MoS_2$ in the promoter region at a sulphur temperature of 250 °C showed the highest monolayer ratio and minimal defects. This study greatly advances our fundamental understanding of TMD growth mechanisms and opens exciting new avenues for tailored synthesis of TMD materials with enhanced properties, improved material controllability and thus reduced device variability.

# Methods

**Substrate preparation**

S1813 (Shipley) diluted by isopropyl alcohol (IPA, Sigma-Aldrich, 99.8%) with a volume ratio of 1:300 is used as the nano promoter. The Si substrate with a 285 nm $SiO_2$ layer (1.5 × 1.5 $cm^2$) first underwent ultrasonic cleaning, including immersion in acetone for 5 minutes, followed by a 5-minute rinse in IPA, and two subsequent rinses with DI water. After the substrate cleaning process, diluted S1813 was spin-coated (by EMS spinner 4000) on half of the sample with an isolation bar at 5000 revolutions per minute (rpm) for 45 seconds, followed by baking at 115 °C for 75 seconds in ambient. The 2-mm wide polydimethylsiloxane (PDMS, Gel-Pak) bar was attached to the middle of the substrate, acting as the boundary between the two regions and preventing the overflow of polymer from one side to the other during the spinning process.

**CVD growth and transfer of MoS$_2$**

The CVD process was conducted in a furnace (MTI GSL-1100X). In a typical growth, N$_2$ (99.999%) was introduced into the tube inlet as a protective carrier gas. Sulphur powders (MaTecK, 99%, 120 mg) were used as the sulphur source and placed upstream in the tube. MoO$_3$ powders (Afla Aesar, 99.9%, 1 mg) were dispersed within a 1cm region in a quartz crucible which is placed at the centre of the furnace. The substrate is positioned 0.5cm downstream from the MoO$_3$ power and tilted at an angle of 120° relative to the tube axis. The temperature of the MoO$_3$ and the substrate was controlled by a pre-set heating program, which was normally increased from room temperature to 835 °C at a rate of 14 °C/minute, held at 105 °C for 1 h and then at 835 °C for 5 min. The sulphur temperature was controlled separately by a heating belt, which was energised when the furnace temperature reached 765 °C. The N$_2$ flow rate started from 500 sccm, was decreased to 20 sccm when the substrate temperature reached 300 °C and then increased to 500 sccm after the reaction completed.

The MoS$_2$ was transferred using a polymer assisted method[50]. A polymer resist, polymethyl methacrylate (PMMA, A3), was spun onto the MoS$_2$ at 3000 rpm for 40 seconds, followed by a 5-minute hot bake. This process forms a polymer film approximately 200 nm in thickness on the sample's surface. Using a razor, approximately 1 mm of the polymer along the chip edge was scratched off to ensure that the polymer film did not adhere to the edge of the chip. Subsequently, the sample was immersed in a saturated sodium hydroxide solution (4M NaOH, Sigma-Aldrich) to etch off the $SiO_2$ layer. Once the $SiO_2$ layer was completely dissolved, filter paper was used to retrieve the floating polymer film, which was then rinsed in clean DI water to remove any residual aqueous alkali and neutralize the transferred $MoS_2$ flakes. After washing, the cleaned film was transferred onto the target substrate. The PMMA film was then removed in acetone for 5 minutes, washed with IPA for 5 min and DI water for an additional 5 minutes.

**Characterization**

An olympus BX51M microscope was used to capture the optical images. OC based analysis and statistical analysis were conducted using Python 3.10. The statistical analysis was based on optical images and conducted using Python code. Considering the image resolution, flakes with a minimum size of 1 μm were included in the statistical count. AFM measurements were conducted on the Oxford Asylum system in tapping mode with Budget Sensors Tap 300 Al − G AFM probe. Raman and PL characterizations were performed using a Witec Alpha 300R system under a 100× objective lens (NA = 0.95, spot size of ~0.7 $\mu$m) with a 532 nm excitation wavelength laser. The laser power of 3.2 mW was used for the Raman and PL mapping. The SEM images were obtained by using a Zeiss Ultra SEM with a field emission gun under InLens mode with an aperture of 30 $\mu$m. The EDX data was collected with a Bruker Nano XFlash 5030 detector integrated with the Zeiss Ultra SEM. XPS measurements were conducted using an Omicron EA 125 Energy Analyzer with a monochromated Al K-alpha source at 1486.7 eV. High-resolution core-level component XPS scans were acquired with a pass energy of 20 eV in high magnification mode, with entrance and exit slits of 6 mm and 3 mm, respectively, giving an overall source and instrument resolution of 0.6 eV. XPS spectra were calibrated to the adventitious carbon C 1 s peak at 285.0 eV. TEM measurements were performed using a FEI Titan 80-300 system with an acceleration voltage of 200 kV. Thermogravimetric analysis (TGA) measurement was carried out using a Stanton Redcroft STA 1500 (Thorn Scientific, UK) calorimeter with platinum-rhodium crucibles and alumina as a reference. The experiments were carried out with a heating rate of 10 °C/min in $N_2$ atmosphere.

# Data availability

The data that support the findings of this study are available from the corresponding authors upon reasonable request. Source data are provided with this paper.

# Code availability

Custom code used in this study is available from the corresponding authors upon reasonable request.

# References

1. Li MY, Su SK, Wong HSP, Li LJ. How 2D Semiconductors Could Extend Moore's Law. *Nature* **567**, 169-170 (2019).
2. Liu Y, Duan X, Shin HJ, Park S, Huang Y, Duan X. Promises and Prospects of Two-dimensional Transistors. *Nature* **591**, 43-53 (2021).
3. Mak KF, Lee C, Hone J, Shan J, Heinz TF. Atomically Thin $MoS_2$: A New Direct-Gap Semiconductor. *Physical Review Letters* **105**, (2010).
4. Radisavljevic B, Radenovic A, Brivio J, Giacometti V, Kis A. Single-layer $MoS_2$ Transistors. *Nat Nanotechnol* **6**, 147-150 (2011).
5. Jadwiszczak J, *et al.* Oxide-mediated Recovery of Field-effect Mobility in Plasma-treated $MoS_2$. *Sci Adv* **4**, eaao5031 (2018).

6. Jayachandran D, *et al.* A Low-power Biomimetic Collision Detector Based on an In-memory Molybdenum Disulfide Photodetector. *Nature Electronics* **3**, 646-655 (2020).
7. Jadwiszczak J, *et al.* Photoresponsivity Enhancement in Monolayer $MoS_2$ by Rapid $O_2$: Ar Plasma Treatment. *Applied Physics Letters* **114**, 091103 (2019).
8. Jadwiszczak J, *et al.* $MoS_2$ Memtransistors Fabricated by Localized Helium Ion Beam Irradiation. *ACS Nano* **13**, 14262-14273 (2019).
9. Jiang J, *et al.* 2D $MoS_2$ Neuromorphic Devices for Brain-Like Computational Systems. *Small* **13**, (2017).
10. Mohapatra PK, Ranganathan K, Ismach A. Selective Area Growth and Transfer of High Optical Quality $MoS_2$ Layers. *Advanced Materials Interfaces* **7**, (2020).
11. Xia Y, *et al.* 12-inch Growth of Uniform $MoS_2$ Monolayer for Integrated Circuit Manufacture. *Nat Mater* **22**, 1324-1331 (2023).
12. Yu H, *et al.* Wafer-Scale Growth and Transfer of Highly-Oriented Monolayer $MoS_2$ Continuous Films. *ACS Nano* **11**, 12001-12007 (2017).
13. Li T, *et al.* Epitaxial Growth of Wafer-scale Molybdenum Disulfide Semiconductor Single Crystals on Sapphire. *Nat Nanotechnol* **16**, 1201-1207 (2021).
14. Yu H, *et al.* Eight In. Wafer-Scale Epitaxial Monolayer $MoS_2$. *Adv Mater* **36**, e2402855 (2024).
15. Jiang H, *et al.* Two-dimensional Czochralski growth of single-crystal $MoS_2$. *Nat Mater*, 1-9 (2025).
16. Singh A, Sharma M, Singh R. NaCl-Assisted CVD Growth of Large-Area High-Quality Trilayer $MoS_2$ and the Role of the Concentration Boundary Layer. *Crystal Growth & Design* **21**, 4940-4946 (2021).
17. Kim H, Ovchinnikov D, Deiana D, Unuchek D, Kis A. Suppressing Nucleation in Metal-Organic Chemical Vapor Deposition of $MoS_2$ Monolayers by Alkali Metal Halides. *Nano Lett* **17**, 5056-5063 (2017).
18. Kim H, *et al.* Role of Alkali Metal Promoter in Enhancing Lateral Growth of Monolayer Transition Metal Dichalcogenides. *Nanotechnology* **28**, 36LT01 (2017).
19. Li S, *et al.* Growth Mechanism and Atomic Structure of Group-IIA Compound-promoted CVD-synthesized Monolayer Transition Metal Dichalcogenides. *Nanoscale* **13**, 13030-13041 (2021).
20. Zhang K, *et al.* Considerations for Utilizing Sodium Chloride in Epitaxial Molybdenum Disulfide. *ACS Appl Mater Interfaces* **10**, 40831-40837 (2018).
21. Lee YH, Zhang, X.Q., Zhang, W., Chang, M.T., Lin, C.T., Chang, K.D., Yu, Y.C., Wang, J.T.W., Chang, C.S., Li, L.J. and Lin, T.W. Synthesis of Large-area $MoS_2$ Atomic Layers with Chemical Vapor Deposition. *Advanced materials* **24**, 2320-2325 (2012).
22. Lee YH, *et al.* Synthesis and Transfer of Single-layer Transition Metal Disulfides on Diverse Surfaces. *Nano Lett* **13**, 1852-1857 (2013).
23. Ling X, *et al.* Role of the Seeding Promoter in $MoS_2$ Growth by Chemical Vapor Deposition. *Nano Lett* **14**, 464-472 (2014).
24. Yang H, Giri A, Moon S, Shin S, Myoung J-M, Jeong U. Highly Scalable Synthesis of $MoS_2$ Thin Films with Precise Thickness Control via Polymer-Assisted Deposition. *Chemistry of Materials* **29**, 5772-5776 (2017).
25. Campbell SA. The Science and Engenering of Microelectronic Fabrication. (2001).
26. Hongzhou Z, Lulin W, Yue S, Kaushik K. Python code for statistical analysis: https://github.com/Lulin-66/MS-Data-analysis.) (2024).

27. Li H, *et al.* From Bulk to Monolayer MoS$_2$: Evolution of Raman Scattering. *Advanced Functional Materials* **22**, 1385-1390 (2012).
28. Yu Y, Li C, Liu Y, Su L, Zhang Y, Cao L. Controlled Scalable Synthesis of Uniform, High-quality Monolayer and Few-layer MoS$_2$ Films. *Sci Rep* **3**, 1866 (2013).
29. Jeon J, *et al.* Layer-controlled CVD Growth of Large-area Two-dimensional MoS$_2$ Films. *Nanoscale* **7**, 1688-1695 (2015).
30. Carvalho BR, Malard LM, Alves JM, Fantini C, Pimenta MA. Symmetry-dependent Exciton-phonon Coupling in 2D and Bulk MoS$_2$ Observed by Resonance Raman Scattering. *Phys Rev Lett* **114**, 136403 (2015).
31. Schaefer CM, *et al.* Carbon Incorporation in MOCVD of MoS$_2$ Thin Films Grown from an Organosulfide Precursor. *Chemistry of Materials* **33**, 4474-4487 (2021).
32. Mignuzzi S, *et al.* Effect of Disorder on Raman Scattering of Single-layer MoS$_2$. *Physical Review B* **91**, (2015).
33. Kaasbjerg K, Thygesen KS, Jacobsen KW. Phonon-limited Mobility in N-type Single-layer MoS$_2$ from First Principles. *Physical Review B* **85**, (2012).
34. Backes C, *et al.* Edge and Confinement Effects Allow in Situ Measurement of Size and Thickness of Liquid-exfoliated Nanosheets. *Nat Commun* **5**, 4576 (2014).
35. Jadkar V, *et al.* Synthesis of Orthorhombic-molybdenum Trioxide (α-MoO$_3$ Thin Films by Hot Wire-CVD and Investigations of Its Humidity Sensing Properties. *Journal of Materials Science: Materials in Electronics* **28**, 15790-15796 (2017).
36. Kumar Singh Patel S, *et al.* Synthesis of α-MoO$_3$ Nanofibers for Enhanced Field-emission Properties. *Advanced Materials Letters* **9**, 585-589 (2018).
37. Feng N, *et al.* A Polymer-direct-intercalation Strategy for MoS$_2$/carbon-derived Heteroaerogels with Ultrahigh Pseudocapacitance. *Nat Commun* **10**, 1372 (2019).
38. Grey LH, Nie H-Y, Biesinger MC. Defining the Nature of Adventitious Carbon and Improving Its Merit as a Charge Correction Reference for XPS. *Applied Surface Science* **653**, (2024).
39. Liang T, *et al.* Intrinsically Substitutional Carbon Doping in CVD-Grown Monolayer MoS$_2$ and the Band Structure Modulation. *ACS Applied Electronic Materials* **2**, 1055-1064 (2020).
40. Geng DZ, X.; Li, L.; Song, P.; Tian, B.; Liu, W.; Chen, J.;Shi, D.; Lin, M.; Zhou, W.; Loh, K. P. . Controlled Growth of Ultrathin Mo$_2$C Superconducting Crystals on Liquid Cu Surface. *2D Mater* **4**, (2016).
41. Zhi Q, *et al.* Dithiine-linked Metalphthalocyanine Framework with Undulated Layers for Highly Efficient and Stable H$_2$O$_2$ Electroproduction. *Nat Commun* **15**, 678 (2024).
42. Gaur AP, Sahoo S, Ahmadi M, Dash SP, Guinel MJ, Katiyar RS. Surface Energy Engineering for Tunable Wettability Through Controlled Synthesis of MoS$_2$. *Nano Lett* **14**, 4314-4321 (2014).
43. Li Y, *et al.* Site-Specific Positioning and Patterning of MoS$_2$ Monolayers: The Role of Au Seeding. *ACS Nano* **12**, 8970-8976 (2018).
44. Govind Rajan A, Warner, J.H., Blankschtein, D. and Strano, M.S. Generalized Mechanistic Model for the Chemical Vapor Deposition of 2D Transition Metal Dichalcogenide Monolayers. *ACS nano* **10(4)**, 4330-4344 (2016).
45. Cao D, Shen, T., Liang, P., Chen, X. and Shu, H. Role of Chemical Potential in Flake Shape and Edge Properties of Monolayer MoS$_2$. *The Journal of Physical Chemistry C* **119.8**, 4294-4301 (2015).

46. Lee C YH, Brus LE, Heinz TF, Hone J, Ryu S. Anomalous lattice vibrations of single and few-layer MoS$_2$. *ACS Nano* **4**, 2695-2700 (2010).
47. Nan HaW, Zilu and Wang, Wenhui and Liang, Zheng and Lu, Yan and Chen, Qian and He, Daowei and Tan, Pingheng and Miao, Feng and Wang, Xinran and others. Strong Photoluminescence Enhancement of MoS$_2$ through Defect Engineering and Oxygen Bonding. *ACS nano* **8**, 5738--5745 (2014).
48. Western SS. X-ray Photoelectron Spectroscopy (XPS) Reference Pages. https://www.xpsfitting.com/2009/10/molybdenum.html.) (2009).
49. M.Ohring. Materials Science of Thin Films (2001).
50. Bie YQ, *et al.* Site-specific Transfer-printing of Individual Graphene Microscale Patterns to Arbitrary Surfaces. *Adv Mater* **23**, 3938-3943 (2011).

# Acknowledgement

This publication has emanated from research conducted with the financial support of Taighde Éireann – Research Ireland under Grant number EPSPG/2020/81 and 20/FFP-P/8727. We are pleased to acknowledge the financial support from Intel Ireland. We also extend our thanks to the staffs of the Advanced Microscopy Laboratory (AML) and the Centre for Research on Adaptive Nanostructures and Nanodevices (CRANN) at Trinity College Dublin for their continued technical support.

# Author contributions

L.L.W., Y.S. and K.K. performed the material growth process. L.L.W. and Y.S. carried out the statistical analysis. Raman and PL measurement were conducted by L.L.W., Y.S. and K.K., with assistance from Y.B.Z., while L.L.W. and Y.S analysed the Raman and PL data. AFM measurements were performed by L.L.W. and Y.S., and SEM imaging and EDX measurement were carried out by L.L.W. and H.Z.W. XPS measurement were conducted by L.G. and C.M.G., end TEM measurements were performed by X.Y.G. and V.N. A.R. conducted the TGA measurements. H.Z.Z. and K.G. contributed to the 3D schematic and illustrations. H.Z.Z. developed the Python code with input from L.L.W., Y.S. and K.K. Data analysis was performed collaboratively by L.L.W., Y.S., K.K., N.M., L.G., X.Y.G., and H.Z.Z. The manuscript was written and edited by L.L.W., H.Z.Z., and P.G., with contributions from all authors. H.Z.Z. conceived and supervised the project. All authors discussed the results and provided feedback on the manuscript.

## Competing interests

The authors declare no competing interests.